\documentstyle{article}

\begin{document}

{\large A PROPOSAL FOR THE QUANTUM THEORY OF GRAVITY}

\bigskip

by Louis Crane, Math Department, KSU.

\bigskip

1. Introduction.

\bigskip

The purpose of this paper is to give a proposal for a candidate for
the quantum theory of general relativity in 3+1 dimensions. The
proposal is easily generalized to a large family of similar theories
which could be interpreted as including various matter fields.
The idea is to modify the 4d topological quantum field theory (TQFT)
constructed in [1] in such a way as
to preserve diffeomorphism invariance, but to introduce propagating
modes. The generalizations would amount to replacing the tensor
category at the heart of the construction in [1] (or equivalently, the
quantum group) by a larger one.

Since the construction in [1] is believed to be a discrete
construction of the $B \wedge F$ theory, while the lagrangian of
general relativity can be written as $B \wedge F$ plus a lagrange
multiplier term [2], it is not implausible to try to quantize general
relativity by augmenting the discrete state sum in [1] with an extra
insertion. We will not try to motivate the insertion by a heuristic
argument involving the two lagrangians, but rather to guess it from
the category-theoretical character of the state sum.

The proposal is, at this stage, a guess. Demonstrating that it is in
fact related to general relativity will require extensive computation
within the proposed model.

This proposal is, in a reasonable sense, a completion of the program
of the paper ``Clock and Category''[3].
We will make the argument that a certain structure in the braided
monoidal categories has the effect of introducing a clock into the
TQFT of [1]. This proposal differs from the proposal of [3], however,
in using an actual modification of the state sum formula of [1],
rather than a simple reinterpretation of it, or a more complex TQFT.
For this change of
direction, I am indebted to Fotini Marcopoulo and Lee Smolin [4], who
recently reproduced  the theory in [1] in a different form, with the
suggestion that the terms of the state sum should be modified, to
include the distinction of the timelike directions.

It was with some reluctance that I refrained from entitling this paper
``what makes a category tick?'' The suggested clock operator is a very
natural piece of the structure of the family of categories from which
TQFTs can be constructed. In this sense, this proposal attempts to
bring quantized general relativity into the sphere of ``categorical physics''.

It should be remembered that the type of tensor categories utilized in
[1] were first introduced into physics in [5] as part of an attempt to
construct string theories. They are ``stringy'' categories in the
sense that their categorical structure is related to operations on the
world sheet of the string. It is possible to regard the model in this
paper as a construction of a sort of string field theory. The
generalizations mentioned above correspond to various choices of
string theories.Whether the anomaly cancellations in string theory
will play a part in the development of this model remains to be seen.

We warn the reader that the treatment in this announcement is not
completely self
contained, but requires a knowledge of [1] and [3], although we do
summarize briefly.

One issue we do not address carefully in this paper is the ``Wick
rotation'' into lorentzian signature spacetime. We believe that this
can be done rather straightforwardly by raising the amplitudes to the
power i, since $e^{is} =(e^s)^i$. Something along these lines does
indeed work in dimension 3 [6].

It is necessary, but very difficult, to address the relationship
between this proposal and the idea of spin network states [6] for the
Ashtekar variables. Spin
networks can be dualized to represent labellings on triangulated
surfaces, so the states in that picture can be related to terms im the
state sum we are using. In our picture, labels that do not sit on
surfaces we choose to treat as observers can be summed over, so that
states on two triangulations coincide if one is a refinement of the
other. If this identification is not used, the spin network states in
the  Rovelli-Smolin picture become problematically multiple. 

As far as the law of propagation is concerned, we have made a guess,
based on mathematical elegance. All the other proposals really amount
to guesses too, related to Hamiltonian operators at boundaries, clock
fields, and various other physical approximations. We may well be in a
period like the origin of quantum mechanics, where apparently very
different proposals turn out to be equivalent, much like wave
mechanics and matrix mechanics. The proposal we are considering has
the advantage of mathematical rigor and computability.

\bigskip

2. State Sums and Tensor Categories.

\bigskip

Let us recall the formula with which we constructed a 4d TQFT in [1]. 
The procedure in that paper is to first label each 2-simplex of a
triangulated oriented 4-manifold with a tilting module of the quantum group
$U_q(sl(2))$ for q a suitable root of unity. More simply, the labels
are half integers up to some finite bound, which should be thought of
as quantum spins. We then put in tensor operators connecting the four
spins at each tetrahedron, and combine the spins and operators around each
4-simplex into a number called a q-15j symbol. The 15 comes from the
fact that a 4-valent tensor operator in this category is given by a
quantum spin, so the expression has 10+5=15 spins in it. We then form
the expression:

\bigskip

\[ \sum \hspace{1ex}
N^{\# vertices - \# edges} \prod_{faces} dim_q(j)
\prod_{tetrahedra} dim_q^{-1}(p) \prod_{4-simplices} 15J_q \hspace{.2in}(*) \]

\bigskip

where $dim_q$ is the quantum dimension, and N is the sum of the
quantum dimensions of the irreducible representations. This expression is independent
of the triangulation, and can be used to construct a 4d TQFT. The
expression should be thought of as the discrete analog of a path
integral.

We can represent this procedure graphically by embedding a certain
surface in the boundary of each 4-simplex. The surface is a union of
tubes dual to each 2-simplex and one per 3-simplex in the boundary of the
4-simplex, joined together to make a closed surface. (The summation in
each 4-simplex is then equivalent to a sum over a basis for the space of conformal blocks
for the surface in the sense of [5].) The surfaces in the boundaries of
adjacent 4-simplices share the tubes corresponding to shared faces and
tetrahedra, so the summations for different 4-simplices overlap.   

The tubular diagrams we use can also be thought of a ribbon graphs
[7]. Remember that a closed ribbon graph can be evaluated to give a
number, but that an open ribbon graph is an intertwining map between
the tensor product of the representations on the open ends at the top
and the same at the bottom. In [1] we used only the numbers on closed
graphs, below, we use the intertwining maps as well.

There is a natural generalization, stated in [1], which uses a more
complex tensor category with analogous structure to the representation
category of the quantum group. The whole proof is very categorical, in
that the fundamental structures of the category translate nicely into
the ingredients of the proof of topological invariance.

This construction can be thought of as filling up the spacetime with a
dense collection of tubes related to the triangulation. We are then
summing over all the spaces of conformal blocks of the tubes, if we think of
the tensor category as the modular tensor category of a rational
conformal field theory as in [5]. It is then a sort of algebraic
miracle that the propagation is independent of the triangulation.

In [1] we considered as spaces of states only formal sums of labels
for representations. This is because all the operators used to
construct the TQFT are intertwining operators, hence acting as a
scalar on each irreducible representation. The operator we introduce
below as a clock does not have this property, so we will be
considering larger spaces of states corresponding to vectors in the
tensor product of the actual representations themselves. Since our
proof of invariance under change of triangulation in [1] is a proof of the
invariance of an operator, we can speak of topological propagation for
this new and larger class of states as well.

A formal heuristic argument can be made which suggests that this theory
is a discrete construction of $B\wedge F$ theory [2]. An important
point of the heuristic is that the representations in the diagram come
from summing over a group element which corresponds to the holonomy of
the connection along some path. Thus in some sense, our state sum has
the right variables to be a discretized quantum theory of gravity. We
do not alter the variables below when we add the ticks. 

The space of physical states of this theory in a closed 3-manifold is
one dimensional, but it is possible to obtain larger spaces of states
associated to 3-manifolds with boundary.  In the program of [3], these
are interpreted as spaces of states which an observer on the boundary 
surface would observe. Thus it is possible to find large spaces of
states in this theory. Unfortunately, in a certain sense, they are
``nonpropagating.'' If we take the cartesian product of the 3 manifold
with boundary by an interval, we get a 4-manifold with corners, which
we can think of as a process of time passing in our theory, this
process gives us the identity operator on all our states.

This is the fundamental difference between TQFT and quantum gravity.
The finite dimensional hilbert spaces in a TQFT can easily be combined
into an infinite dimensional limit space [9], but the non-propagation is a
basic problem. On the other hand, diffeomorphism invariance is a
common feature which we would like to preserve. Finally, the TQFT we
have outlined has no timelike directions. 

So can we modify the construction in [3] to solve all these problems?

\bigskip

3. The tick of time

\bigskip

At this point, we explain the modification of the state sum in [1]
which could transform the resulting theory into general
relativity. Our discussion as we have warned the reader, is not self 
contained, but presupposes a
knowledge of [1]. We also do not supply the proof of isotopy invariance, which is not
difficult. 

We propose to place insertions of certain operators on the category of
representations we used in [1] inside the diagrams corresponding to
the triangulated  4-manifold in such a way as to introduce a clock
into the theory. The operators to be inserted are the canonical grouplike
element of the quantum group $U_q(SU(2))$ for which the labels on the
faces in the TQFT of [1] are representations. This hopf algebra has a
canonical grouplike element and all other grouplikes are powers of it; the larger tensor
categories would have different possible choices for a grouplike on them. For
a discussion of the grouplike element and its properties see [10,11].

The grouplike in $U_q(SU(2))$ is easy to understand. It is $q^H$,
where H is the q analog of the diagonal matrix $J_z$. The grouplike
property is related to the fact that $J_z$ remains a good quantum
number for q-deformed spins. ( $J_x $ and $J_y$ do not.)

A grouplike element in a hopf algebra obeys the equation $ \Delta g= g
\otimes g$ , $ \epsilon(g)=1, s(g)=g^{-1}$ . The first equation means that when applied to a representation which
then has a tensor operator taking it to a product of two
representations, it can be ``passed through'' the vertex, to act on
each of the two representations below the vertex. The second implies
that when we change a downward tube to an upward one in a diagram the
grouplike switches to its inverse. These properties allow
us to use the grouplike element to put an insertion into the state sum
on a hypersurface in such a way that the result is unaffected if the
hypersurface is isotoped. The hypersurface cannot be isotoped to the boundary.

We refer to a hypersurface on which the tubes are decorated as a
``tick,'' which we think of as a space slice where the clock time
advances by one Planck time. The model would be to calculate the state
sum for the TQFT with many ticks.

In order to define the tick precisely, it is necessary to pick an
ordering of the set of tetrahedra in the triangulation. This gives a
direction on each tube in the model (which is dual to a face in the
boundary of a particular 4-simplex). We next pick a hypersurface in
general position with respect to the triangulation, which traverses
the cone on each tetrahedron with apex the center of each 4-simplex
which the tetrahedron bounds regularly and once only or not at all. We
further require the hypersurface to be oriented with a collar
neighborhood.

For the nonmathematician, we are requiring the hypersurface to divide
the spacetime into two sides, a past and a future.

The exact prescription is now to decorate the tubes dual to the faces
of each tetrahedron with copies of the grouplike if the tetrahedron is
ordered lower than the one on the other side of the face and the
center of the 4-simplex is to the past of the tetrahedron, reversing to
the inverse of the grouplike if the tetrahedron is ordered higher than
its opposite or if the center of the 4-simplex is in the future, and
labelling with the grouplike if the opposite tetrahedron is lower and
the center is in the future of the tetrahedron. It is equivalent to do
the labellings only on faces which bound between traversed and
nontraversed tetrahedra (always in the boundary of a particular
4-simplex), because of cancellations.

Basically, this prescription works because a spin depicted as a down
line is expanded in a dual basis from an up line, so that the
grouplike is transformed to its inverse.

A mathematical paper with the details and proof of isotopy invariance
of this prescription will follow [12]. It is also possible, but
unenlightening to include ``wigglier'' hypersurfaces.

The idea now is to put a long sequence of ``ticks'' in a 4-manifold with
corners, and to see how the states on neighboring surfaces interact. It is not
necessary for the string of hypersurfaces to be parallel, they can
also spiral or have various topological configurations.

Let us mention that the 4d state sum we are modifying, and
hence the TQFT and also the ticking model we construct from it, are
actually more flexible than our description shows. The ``blob lemma''
of [1] shows that we can replace the triangulation by any polyhedral
decomposition. Furthermore, the work of Roberts [13], shows that we
can replace the triangulation with a complicated weave in the boundary
of a 4-ball. The tick is equally well defined in these situations.

Thus we can represent our ticking model as a weave passing through a
long series of ticking surfaces. This may make calculations easier.

We believe that the procedure we outline here has several properties
which are suggestive of general relativity. First, the independence of
the tick location from the triangulation is reminiscent of the
difference between clock time and coordinate time. Second is the
manifest diffeomorphism invariance of the constructed theory. Third is the
natural introduction of a fundamental length scale into the theory,
while still being able to consider any triangulation. Fourth is the
identification of the states in this theory with the spin network
states derived from the loop variable picture [4].

None of this shows that the ``ticking'' theory we construct actually
resembles general relativity in any way. That can only be addressed by
actual calculations in the ticking model. It is possible to say, at
least, that the model is well defined; so that a number of
calculational investigations are possible. Let us list some natural
things to investigate: 1. the analog of the Wheeler -De Witt equation
at a tick; 2. the behavior of sectional curvatures around an edge near
a tick; 3. the propagation of states through a sequence of
ticks.

\bigskip

4. Towards an investigation of the model.

\bigskip

At this point we wish to make some preliminary remarks concerning the
physical interpretation of the model. 

First, we state without proof some facts
about the effect of insertion of the grouplike in a diagram in the
category.

Insertion of a grouplike on a single strand of a closed diagram has
the effect of multiplying the evaluation of the diagram by the quantum
dimension of the representation on the edge where the grouplike is
inserted. Insertion of grouplikes at more than one place, however, has
a more complex effect, since the operations are correlated. If two
edges are joined to a third at a vertex, the joint effect of
grouplikes on both of them is equal to the effect of a grouplike on
the third. This means that states on the pair which are strongly
correlated (ie add up to a larger representation in the tensor
product) are weighted more heavily by the effect of the insertions of
the grouplikes.

The definition of a ``tick'' we made above reduces to putting
grouplikes on the tubes dual to the faces of one tetrahedron at a
time.

Since each set of four such tubes is connected at the center of the
tetrahedron in our model, the effect of a tick will be to cause the
states attached to the faces of a tetrahedron to become correlated. We
believe that this has a classical limit which reproduces the form of
Einstein's equation in Regge calculus. If this is so, then our model
will genuinely have general relativity as a classical limit. The
motivation for this idea is to think of the minimal example, in which
the representations are q-deformed spins, and the quantum dimension is
the q-analog of angular momentum. The spin states should be thought of as
generalized  fourier coefficients of sectional curvatures. The correlation between
them would correspond to some weighted sum of the sectional
curvatures at a tetrahedron, which is a discetized analog of the Ricci
curvature. A more refined analysis of this argument, with especial
care taken as to the effects of Wick rotating into the lorentzian
regime will be necessary to see if there is really any validity to it.

In general, since each tick is producing correlations between
neighboring observers, we expect that our model will show propagation.
Also, different vectors in the same irreducible representation will
now have fundamentally different evolutions because of the tick.  

Now we can understand the problem of writing down propagating modes.
The eigenstates of $J_z$ are multiplied by appropriate powers of q as
we pass a tick. This is analogous to the behavior of eigenstates of
the momentum operator for a quantum mechanical free particle. in order
to see that quantum mechanical particles propagate, we need to form
wave packets.

We can imagine joining together all the tubes labelled on a tick to a
single tube, which would end up labelled by a very large
representation which would be the tensor product of those on the
individual tubes. The $J_z$ operator for the combined representation
would be like a Hamiltonian for the total system. The fact that $J_z$
is a grouplike allows us to break it up into local pieces.

The problem of actually constructing a propagating mode will use this
local decomposition of the tick/ Hamiltonian. Interpretation is made
more subtle by the diffeomorphism invariance of the construction,
which must be somehow broken in a semiclassical picture. The necessary
ingredients all seem to be present, however.

In
order to actually demonstrate propagating modes, we would need to pick
a background spacetime geometry (determined by choosing the
irreducible representations at the 2-simplices rather than summing
over them) and studying how
particular vectors in it propagate in the state sum with ticks. It
should be possible to construct wave packets using the different
phases different vectors pick up at the ticks. This will be a
formidable computation, however.

The passage to our larger notion of states, mentioned above, is
crucial to allow this. The states of the TQFT, which do not
distinguish vectors inside the representations, do not propagate under
the tick.

Assuming this proposal goes through, the propagating states would live
on 2-simplices of the triangulation. This suggests that they would
look a lot like gravitons, i.e. have a tensorial character.

\bigskip

5 Strings and things.

\bigskip

It is natural to ask at this point what connections this picture may
have with string theory. The most salient connection is the fact that
both the 4d-TQFTs we modify here and rational conformal field theories
are constructed from the same family of tensor categories [5]. The
same categories are also used to construct the state spaces on
surfaces. (We think of surfaces in spacelike 3-manifolds as the
boundaries where observers observe. See [3].)

It is worth recalling that the program of string theory faltered
through its inability to construct a ``string field theory'' to which
the string amplitudes gave perturbative terms. The construction we are
proposing amounts to filling the spacetime with a network of Planck
scale strings, obtaining a propagation law unaltered by any further
refinement of the network. This circumvents the nasty mathematical
problems of constructing a string field theory on loop space.

Also, the stringy states on the tubes which meet the past and future
boundaries of our spacetime are combined to produce the vector spaces
on surfaces which we interpret as relative states for observers.
Perhaps, since the state space of an observer is dominated by states
which almost turn it into a black hole, the strings really come into
the theory as boundaries for quantum observers which are rather like
black holes. In other words, maybe strings are things.

The most difficult part of our model to interpret in stringy terms is
the grouplike in the hopf algebra. Here we note that Moore and Seiberg
in [5] actually regarded a group as a ``classical'' rational conformal
field theory. in that analogy, elements of the group are ``grouplike''
in the same sense as the grouplike element of the quantum group. The
grouplike represents a remaining symmetry unbroken in the deformation
producing the quantum group.

\bigskip

6. Conclusions.

\bigskip

At this point our proposal is really only the beginning of a program
of investigation. It is certainly natural to ask if any other 
modifications of the state sums of [1] exist which might be worth
studying.

One proposal would be to modify the 15j symbols by cutting them along
the hypersurface and inserting some carefully chosen spin diagrams.
One possibility would be to connect adjacent tubes with crosstubes
labelled with the basic representation. This is strongly reminiscent
of the hamiltonian proposed for quantum gravity in [14].

A drawback of such a procedure is the difficulty of finding one with
any sort of invariance property. This could be a simple lack of
imagination, but the elegance of the behavior of the grouplike is seductive.
Also, any such proposal would not distinguish vectors inside a representation.

A drawback of this proposal is the absence of any heuristic argument
connecting the laplace multiplier term in the lagrangian of GR to our
picture. Perhaps more careful thought would suggest a different
insertion from that quarter.

At any rate we have proposed a model with the following features:

\smallskip

1. A natural distance scale

\smallskip

2. Variables which are reasonable for a discretized version of general
relativity

\smallskip

3. a perturbation of $B \wedge F$ theory

\smallskip

4 diffeomorphism invariance

\smallskip

5. a clock

\smallskip

6. (plausibly) propagating modes resembling gravitons.

\smallskip

Clearly, this model deserves further study.

Let us remind ourselves of another feature of string theory, namely
the cancellation of conformal anomalies and the related choice of a
special class of models. This phenomenon has a counterpart in the
construction of 3d TQFTs from modular tensor categories: the
amplitudes for propagation of topological states has a phase ambiguity
unless we choose a model with ``central charge'' a multiple of 24
[15].

We have not yet computed the propagation of states on surfaces
carefully enough for the ticking model, but it is very likely that a
phenomenon of this type will recur. If so, there will be a topological
reason to prefer the same models which appeared as string motivated
GUT theories durring the heyday of the string. If this does turn out
to be the case, the physical applicability of the ticking models will
be more direct.

As we mentioned in the introduction, there is a natural proposal for
Wick rotation for this model, namely raising each term in a state sum
to the power i. It needs to be carefully studied if the ticking
together with this prescription makes the timelike directions tend to
align.

Another interesting feature of this model is that a tick on a closed
contractible hypersurface can be contracted away, leaving no result.
It is tempting to try to use this feature of the ticking model to shed
some light on the difficult questions of time evolution of states
which fall into black holes. The suggestion is that classical states
with event horizons may occur in quantum evolutions which do not
possess them. What this may tell us about information loss remains to
be explored.

In this paper we have been constructing the ``stringy'' tensor
categories as representations of quantum groups. They also appear as
representations of loop algebras with central extensions. If we could find a natural choice of
grouplike for the universal enveloping algebras of the Kac Moody
algebras, then we could repeat our construction, with genuinely
stringy state spaces on surfaces. The grouplike would have to behave
well with respect to the tensor product of Moore and Seiberg [5]. This
would lead to rather different problems of physical interpretation.

\bigskip

{\bf Acknowledgements}

\bigskip

This paper was strongly influenced by conversations with Fotini
Marcopoulo and Lee Smolin. The precise expression for the tick was the
result of conversations with David Yetter.

\bigskip

ADDENDUM

\bigskip

We wish to note that the definition of the tick above is actually more
flexible than we realized. It is possible to modify the construction
so that the grouplike appears in the same places as in I, but raised
to the power of the absolute value of the casimir of the
representation on the tube it is attached to. This will have the
effect that if a vector in a representation on some face of the
triangulation  is followed through the
tubular diagrams in the 4-manifold to another face, the contributions
along various paths will pick up different phases, corresponding to
the total ``lengths'' of the paths. This seems much like a discrete
analog of the path integral expression for a propagator for a massless
field in a gravitational background. 

We believe that with this modification the model has every indication
of possessing propagating modes, although demonstrating them will
still be a formidable calculational problem.

\bigskip

{\bf REFERENCES}

\smallskip

1 L. Crane and D. N. Yetter, A Categorical Construction of 4d
Topological Quantum field Theories, in Quantum Topology, L. H.
Kauffman and R. A. Baadhio, eds World Scientific 1993 120

see also:

L. Crane, L. H. Kauffman and D. N. Yetter, State Sum Invariants of
4-Manifolds, J. Knot Th. and Ram. to appear

\smallskip

\smallskip

[2] J. C. Baez, degenerate Solutions of General Relativity from
Topological Field Theory gr-qc 9702051

\smallskip

[3].L. Crane, Clock and Category; is Quantum Gravity Algebraic? J. Math.
Phys. 36 (1995) 6180

\smallskip

[4] F. Markopoulou and L. Smolin, Causal Evolution of Spin Networks.
gr-qc 9702025

\smallskip

[5] G. Moore and N. Seiberg,  Classical and Quantum Conformal Field
Theory, Commun Math. Phys.  123 (1989) 177-254.

\smallskip

[6] J. W. Barrett and T. J. Foxon, Semiclassical Limits of Simplicial
Quantum Gravity, Class. Quant. Grav. 11 (1994) 543

\smallskip

[7] N. Reshetikhin and V. Turaev, Ribbon Graphs and their Invariants
Derived From Quantum Groups, Comm. Math. Phys. 127(1990) 1-26

\smallskip

[8] H. Ooguri,  Topological Lattice Models in Four Dimensions,
Mod. phys. Lett. A7 (1992) 2799

\smallskip

[9] D. Yetter, triangulations and TQFTs, in Quantum Topology,
Kauffman and Baadhio eds, World Scientific, 1993

\smallskip

[10] L. H Kauffman and D. E. Radford, Invariants of 3-dimensional
manifolds derived from finite dimensional hopf algebras J. Knot Th. Ram.
v4 1 March 1995 131-162

\smallskip
 
[11] ibid. A Necessary and Sufficient Condition for a finite dimensional
Drinfeld double to be a ribbon Hopf algebra, J. of Alg v159 1 
(1993) 98-114

\smallskip

[12] L. Crane and D. Yetter, Ticks in TQFTs, in preparation.

\smallskip

[13] J. Roberts, skein Theory and Turaev-Viro Invariants, 1993 preprint

\smallskip

[14] Carlo Rovelli and Lee Smolin, Spin Networks and Quantum Gravity,
gr-qc preprint

\smallskip

[15] L. Crane, 2-d Physics and 3-d Topology, Commun. Math. Phys.
 135 (1991) 615-640.

\smallskip

\end{document}